\documentstyle[twocolumn,prc,aps,epsfig]{revtex}
\input epsf
\newcommand{\be}{\begin{eqnarray}}
\newcommand{\ee}{\end{eqnarray}}
\def\lsim{\mathrel{\rlap{\lower3pt\hbox{\hskip1pt$\sim$}}
     \raise1pt\hbox{$<$}}} 
\def\gsim{\mathrel{\rlap{\lower3pt\hbox{\hskip1pt$\sim$}}
     \raise1pt\hbox{$>$}}} 

\begin{document}

\twocolumn[\hsize\textwidth\columnwidth\hsize\csname 
@twocolumnfalse\endcsname

\title{Can Binary Bound States
in a Strongly Coupled Quark-Gluon Plasma \\
be observed via dileptons and photons?}

\author { Jorge Casalderrey-Solana and
Edward V. Shuryak  }
\address { Department of Physics and Astronomy\\
State University of New York,
     Stony Brook, NY 11794-3800}

\date{\today}
\maketitle
\begin{abstract}
Recently there was a significant change of views on physical
properties
and underlying dynamics of Quark-Gluon Plasma at $T=170-350\, MeV$,
produced in heavy ion collisions at RHIC. Instead of weakly coupled
gas of quasiparticles, it is rather a liquid-like matter with multiple
bound states. In this paper
we discuss how one can test these ideas experimentally,
using the ``penetrating probes'' and looking for certain
 peaks at some invariant masses. In dileptons the most promising
are modified 
$\rho,\omega$,  with  $M(T\approx T_c)\sim .5 \, GeV$ and also near
zero binding
 at $M(T\approx (1.5-2)T_c)=1.5-2\, GeV$.
 We also discuss the observability of
peaks corresponding to scalar/pseudoscalar mesons in the two-photon channel.
\end{abstract}
\vspace{0.1in}
]
\begin{narrowtext}
\newpage

\section{Introduction}

\subsection{The penetrating probes}

To use photons and  dileptons  as ``penetrating probes"
   to study the earlier
stages of heavy ion collisions was  one of the first suggestions
to study quark-gluon plasma (QGP)     \cite{Shu_78}.
Although these particles are produced much less copiously compared to hadrons,
and thus they are more difficult to study,  photons and leptons are
of special importance because they
do not suffer final state interaction and therefore can bring us
direct information about the earliest (the hottest) stages of the
   collision. In this respect one may compare them to e.g. solar neutrinos,
which bring direct information about the conditions at Sun's interior.

   It was  argued in  \cite{Shu_78}
  that because of the QCD phase transition, one should expect
the invariant  mass ($M^2=(p_{l+}+p_{l-})^2$)  spectra 
of penetrating probes produced
in QGP to be qualitatively different from those originating
in hadronic matter.  However only
 monotonously decreasing mass spectra were expected from QGP,
  in contrast to familiar vector meson
 $\rho,\omega,\phi,J/\psi...$ peaks from the hadronic phase.
 The rates were calculated in a picture of
weakly interacting and freely moving near-massless
  quark quasiparticles, below  wQGP for brevity.

  Recently a radically new picture of QGP has been developed,
to be described in next subsection, based on much stronger coupling
between quasiparticles, to be referred to as sQGP. One aspect of it
is  that
the  meson-like  bound states continue to exist at
$T=(1-2)T_c=170-350\, MeV$
(the temperature range corresponding to QGP at RHIC),
although in a strongly modified form. Furthermore, even at
higher
$T$ when there are no bound states, there exist near-threshold
enhancement
which may still be used to infer the value of the quasiparticle masses and
the strength of their interaction.

  As penetrating probes are emitted
during the whole expansion of the fireball,
the  evaluation of observable spectra  
includes two very different steps.
(i) First one evaluates a {\it production rate} in
equilibrium matter at a given $T$, $dR(T)/d^4xd^4q$
per time per volume per 4-momentum. The second
step (ii)
is the space-time integration over the 4-volume of the expanding fireball,
from its creation to final freezeout.
 At the first step (i) there will be certain peaks
at fixed values of  mesons masses, but since they are in general
$T$-dependent, $m_V(T)$, the time integral
at the second step will in general
smoothen them out. Only peaks corresponding
to special points can survive and be observable, as we detail below.

 Let us start with the basic expression for the
dilepton production rate 
\begin{eqnarray}
{dR(T)\over d^4x d^4q} = {\alpha^2 \over 48\pi^4} {1\over e^{q_0\over T}-1}F
\end{eqnarray}
where $\alpha$ is the electromagnetic coupling, $q_0$ is the dilepton
energy
and the ``formfactor'' $F$ for
two well-tested processes are,
\begin{eqnarray}
\label{ff}
F = \cases{F_H \buildrel \rm def \over = {m_\rho^4\over[(m_\rho^2 -
    M^2)^2 + m_\rho^2\Gamma_\rho^2]}, \hspace{.3cm}(\rho)\cr
F_Q \buildrel \rm def \over = 12 \sum_q e^2_q\left(1+{2m_q^2\over
  M^2}\right) \left(1-{4m_q^2\over M^2}\right)^{1\over 2}, (wQGP)\cr}
\end{eqnarray}
where the former one corresponds to
  $\pi\pi\rightarrow\rho\rightarrow l^+l^-$ annihilation in
a vacuum or hadronic phase at small T,
written in standard vector-dominance form.
The latter expression
corresponds to   $\bar q q $ annihilation
similar to partonic Drell-Yan process.
If one ignores quark masses in $F_Q$, it is just a sum
of squared charges for all relevant quarks, $u,d,s$ of all colors.
This basic case we will use as our
``standard candle'' below, normalizing all predictions to
the ``standard wQGP rate'' with  $F_{wQGP}=24$.
Detailed calculation of the final
dilepton spectra for such ``standard candle''
model  were done
by Rapp and Shuryak \cite{RS_IMR} (including specific
acceptance of the SPS NA50 experiment) and for RHIC
by Rapp \cite{Rapp_RHIC}. For a review see \cite{RW}.

In terms of the imaginary part of the photon self energy wQGP limit means
\be
\label{ImPipert}
\Im{\Pi_{em}}=-\frac{M^2}{12\pi} N_c \sum_{q=u,d,s} (e_q^2)
\ee

If quark quasiparticles in the QGP phase are not massless (and current
lattice calculations indicate they are as
heavy as $1 GeV$ in the region of interest $T_c-3T_c$) one would like
to
correct for that using the formulae from above.
 This fact leads to a trivial modification to the previous expression
(that is relevant in the IMR)

\begin{eqnarray}
\label{impi}
\Im\Pi = \cases{  0,\hspace{.3cm} (M^2<4m_q^2) \cr
\frac{-M^2 }{12\pi} N_c  \sum_q e^2_q \left(1+{2m_q^2\over
  M^2}\right) \left(1-{4m_q^2\over M^2}\right)^{1\over 2}, \cr (M^2>4m_q^2)\cr
}
\end{eqnarray}

Furthermore, we expect nontrivial
modification of the annihilation cross section of quarks
into leptons due to the attractive interaction
between them. This modification will be specially
important for near threshold production,
where we will use non relativistic Green\'{}s functions.
Modifications of the rates are also expected due to the presence of bound 
states, where, as we sill see, the problem is
intrinsically relativistic.

   Dileptons
 produced by vector resonances ($\rho,\omega,\phi,J/\psi$)
are described by the first expression in (\ref{ff}),
which can be used both in a hadronic phase and in sQGP.
There exist large literature
on matter-induced modification of mesons in hadronic phase.
For sufficiently dilute matter one can use linear density
 approximation, in which the modification of real and imaginary part
(the mass and the width) of a state is simply
determined by forward scattering amplitudes  {\em in vacuum}
which are often experimentally known.
For example,
$\pi\rho$ and $N\rho$ scattering   used in \cite{Shuryak:1991hb}
are mostly responsible for modification of $\rho$ resonance.
 Quite a number of works following such ideas found
relatively modest  shifts of resonance masses
(e.g.$m_\rho$) downward and some broadening\footnote{As shown in
\protect\cite{Shuryak:2002kd}, mass reduction leads to some suppression 
of the decay, which often tends to cancel widening from a
 rescattering. This was experimentally confirmed by STAR experiment at
 RHIC
\protect\cite{Fachini:2003mc}, which found no width change of the
 $\rho$ observed via $\pi\pi$
at late stages.
}. 
Further assumptions are needed when the matter
is not dilute:
e.g. the Li-Ko-Brown model\cite{LKB} assumed that all hadronic masses
are shifted proportional to
Walecka-type scalar mean field, related to local matter density.
As example of self-consistent approach see e.g. 
a set of coupled gap equations discussed by
Rapp and Wambach  \cite{RW}: it predicted very strong
broadening of the $\rho$ meson
due to mixing between $\rho$ with
excitations of the lowest baryon resonances,
such as\footnote{This notation means a nucleon excited into a
baryonic resonance by vector meson absorption.} $N^*(1520)N^{-1}$.


   Let us now proceed to a brief review of dilepton experiments.
Since these  kind of measurements are generally much more
difficult to do
(there are large background  to be rejected, while the process itself
has  small cross section),  they
have a sad tendency to come ``too late'', right before
the program gets closed. 

 At Berkeley
BEVALAC the DLS dilepton spectrometer has found
a spectacular effect, but the  detector efficiencies  were debated
years after its last run\footnote{There are no official pulication
  yet,
but preliminary results  from HADES experiment at GSI recently
reported have not
confirmed this effect in its first run.}.
Brookhaven AGS program had no dilepton experiments, while
CERN SPS program included:
(i) CERES (NA45), which studied the low mass  (M=0-1. GeV) $e^+e^-$ pairs,
(ii) HELIOS-3, which studied medium mass $\mu^+\mu^-$''  M=1-2 GeV,
and (iii) NA38/50/60 concentrated on  $\mu^+\mu^-$ with M=2-4 GeV.
All of them
observed quite significant enhancement over ``standard hadronic sources'', 
ranging
from factor 5 at CERES (in some kinematic region) to about 3 at
NA38/50 for intermediate masses
\footnote{At high masses region, $M>3$ GeV, the
dilepton production is well described by simple partonic Drell-Yan
process.}
M=2-3 GeV.
The observed
effective ``$\rho$
melting'' at CERES and thermal continuum at NA50
indicates an approach toward chiral symmetry restoration
and QGP\cite{RS_IMR,Kempfer_etal}.
This is quite puzzling,
since at the SPS only a small fraction of the
space-time volume contributing to dilepton production is expected to
originate from QGP, while most of it should still
be from the mixed or hadronic phase.
The NA50 experiment now evolved into  NA60 which is supposed to
resolve
the remaining uncertainty between thermal dileptons and
charm contribution, as well as significantly increase mass resolution
and tell us about possible modifications of narrow $\omega,\phi$ resonances.

The dileptons are supposed to be
the
main part of the RHIC  program: PHENIX was aimed at
that from the beginning and now also STAR detector has electromagnetic
calorimeter. Large-statistics Run-4 data are now being analyzed,
and the only published data so far are
PHENIX  single electron
production \cite{PHENIX_singleelectrons},
which is presumably dominated by charm decays.

Matter modification of more narrow $\omega,\phi$ resonances 
can obviously be observed only for
 a fraction of these particles decaying
inside the fireball. Resonances decaying after freezeout
have the usual vacuum mass, and these decays produce large ``hadronic 
background''
for such measurements: in fact 
so far none of the existing dilepton experiments had sufficient resolution to
observe these modifications.
  However for sQGP vector mesons the mass modification is very large,
e.g. we will discuss below  peaks with masses up to $M\sim 2 \, GeV$,
and so in this case the resolution  is not so much an issue.  

  In principle, the  $\gamma\gamma$ channel of the decays provides
another window into early-time spectroscopy, as photons are
penetrating probes as well.  Still, not much discussion
of those can be found in  literature. 
In-matter modification of the {\em pseudoscalars}
$\pi,\eta$ are expected to be quite small, as
 these masses are protected by chiral symmetry and are not expected to
be modified much in all hadronic phase $T<T_c$. The experimental
 observability
of these shifts in $\gamma\gamma$ channel is next to impossible,
due to huge background from similar decays after freezeout.
Scalars and tensors are considered to be too wide even in matter, to get
any meaningful signal even before in-matter modification.

The situation again should be quite different for $\pi^0,\eta,\eta'$ and 
 scalars originating from  sQGP\footnote{We thank V.Zakharov who 
suggested it to one of us (ES) as a possibility, in a discussion after
his talk.
}, since very large mass shifts
 are expected. We return to their discussion in section
 \ref{sec_gammas}
below.

\subsection{ Bound states in QGP and the penetrating probes }
The quark-gluon plasma (QGP) was defined as a state of matter in which
a color charge is screened \cite{Shu_78} rather than confined.
According to lattice and experimental results it takes place above
the critical temperature $T_c\approx 170\, MeV$.

At high temperatures $T\gg T_c$ it should be
a gas of weakly interacting quasiparticles (modulo long-range
magnetism), wQGP.
However a physical picture of QGP structure at not-too-high $T$,
in the temperature range
$(1-3)\,T_c$, recently underwent a radical change \cite{SZ_rethinking}.
The interaction seems to be
sufficiently strong to produce multiple bound
states
of quark and gluon quasiparticles, and we will
refer to such  matter  as strongly coupled QGP
(sQGP for brevity).

Using lattice-based potentials, it was shown~\cite{BLRS,SZ_bound}
that a picture of binary bound
states may
provides a consistent description of several
previously disconnected lattice observations, such as (i)
bound states for charmonium and some light $\bar q q$ states~\cite{MEM}; 
(ii)
static potentials~\cite{potentials}; (iii) quasiparticle 
masses~\cite{masses};
and (iv) bulk thermodynamics~\cite{THERMO}.
However more complicated structure can also be present,
perhaps in line with a liquid-like behavior  needed to
understand robust
collective flow phenomena at RHIC, well described by ideal hydrodynamics.

Studies of strong coupling and related binary states are also under
way
in other fields of physics.
In particular, the
$\cal N$=4 supersymmetric Yang-Mills
theory at finite $T$ is a perfect toy model, in which
a parametric transition from weak to  strong coupling can be
traced, see discussion of binary states in
\cite{SZ_CFT}. Another strongly-coupled
system is produced experimentally,
by cold trapped atomic gases, by tuning the scattering
length to very large values via Feshbach resonances.
Hydro-type flow  and small damping of oscillations observed for these systems
indicate liquid-like properties of matter in this case as well\cite{GSZ}.

The goal of this paper is to formulate how one can tell the difference
between the
wQGP (with light and weakly interacting quarks) from
the sQGP picture outlined above, by using a dilepton penetrating probe.
More specifically, we will discuss the following questions:\\
(i) Is it possible to measure the interaction between quark
quasiparticles directly, using the dilepton probe?\\
(ii) Where are the most significant
differences
between sQGP and wQGP scenarios, in respect to the dilepton production?\\
(iii) How large are they and whether one has a
chance to observe them, at RHIC and elsewhere?

We identify two such mass regions: \\
(i) $M\sim 1.5-2 \, GeV$ corresponding to $T\approx T$(zero binding) and
the edge of the quasiparticle continuum, at the initial QGP at RHIC.\\
(ii)  $M\sim 0.5 \, GeV$, where we expect
to
see the contributions of  the modified  vector mesons at $T\approx T_c$.

\begin{figure}[t]
\centering
\includegraphics[width=6.cm]{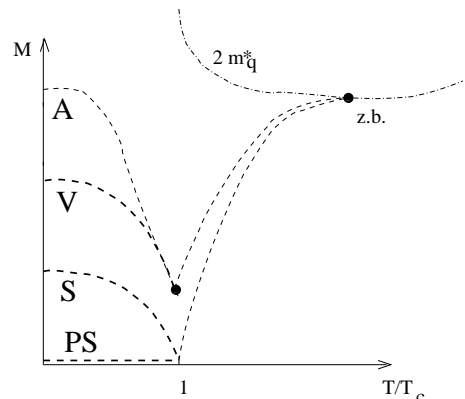}
   \caption{\label{fig_masses}
Schematic  $T$-dependence of the masses of $\bar q q$
states. $A,V,S$ and $PS$ stand for axial, vector, scalar and pseudoscalar
states.
The dash-dotted line shows a behavior of twice the quasiparticle mass.
Two black dots indicate places where we hope the dilepton signal may
be observable.
}
\end{figure}
The basic idea is very simple: the
probability of $\bar q q\rightarrow l^+l^-$ process is
enhanced by the initial state attractive interaction.
Attractive interaction obviously correlate
$\bar q$ and $q$ in space and increases the chances to
find $\bar q q$ close to each other and annihilate.
In such general form, the enhancement persists
whether the potential is deep enough to make bound levels or not,
and whether quark quasiparticles have small or large width.

In the case of pure Coulomb
interaction this phenomenon is
well studied. The relevant parameter  $z=\pi\alpha/v$ contains the
ratio of the coupling strength to velocity,
and the enhancement
is in this case
given by the well known Gamow factor\footnote{The sign in exponent
is the opposite to that on
the original Gamow factor, for alpha-particle interaction
with nuclei or that used for HBT
correlations, corresponded to a $repulsive$ Coulomb force.}
\be
  F_{Gamow}= {z \over 1-exp(-z)}.\ee
Note that the result is $\sim 1/v$ at small
$v$ (large $z$) and 1 at small $z$. It
cancels the velocity in the phase space and makes the
production rate to jump to a $finite$ value at the threshold.

A good example of the QCD-based color-Coulomb enhancement
is the cross section for the production of the top quark pair
$\bar t t$, see  \cite{FK_88,Strassler:1990nw}. The enhancement
calculated in these works
is  found to be quite significant, in spite of the fact
that  rather large width of the top quarks precludes them from forming
topponium bound states. We will use methods developed in these works
in  section IIb.

In section \ref{sec_newqgp} we will discuss the dilepton production
in QCD
at $T>T_c$. But before we get to discuss details, let us now point out
our main ideas. In Fig.\ref{fig_masses} (which is a modified version
of a figure present already in \cite{SZ_rethinking}
we show schematic temperature dependence of lowest $\bar q q$ states.
Dileptons can only come from colorless vector pairs
 (marked by $V$). 

New (sQGP) part of this picture is at $T>T_c$,
which starts and ends at points marked by small black circles.
At any given $T$ there are two peaks in the spectral density,
corresponding to invariant masses of $T,L$ components. On top of that,
there is a threshold bump at $2m^*_q(T)$, which exists even at
$T$ so high that no bound states exists.
Unfortunately, one can only observe a signal integrated over the
expanding fireball, or over all temperatures involved between the
initial and freezeout ones. The integral tends to wipe out
peaks at intermediate locations, unless
there are special reasons for them to survive.
Two of the black dots, at the mixed phase $T=T_c$, have a benefit
of long time spent there during expansion. The same is true for
the third black point, due to near-constancy of the mass of weakly
bound states at $T=1.5-2 T_c$. The near-threshold bump
is at about the same location at higher $T$ as well. Thus one may hope
(and we  will show it
below) that the structures corresponding  to these endpoints (black
circles
at Fig.\ref{fig_masses}) may be detected.

Although at finite
temperatures and non-zero
momentum relative to heat bath
 are split into distinct longitudinal and transverse
($L,T$) modes, those should coincide at zero momentum.
Since all masses are large compared to relevant $T$ at the time,
only pairs with small momentum are actually produced.
Furthermore, lattice experience of similar quasiparticle modes
indicate that they follow, at least approximately,  the
vacuum-like dispersion law $\omega^2=p^2+M^2$, and so using
the invariant mass rather than energy would take care of 
canceling the momentum dependence.

In Fig.\ref{fig_masses} we have shown only states made of $u,d$
quarks,
ignoring the strange one. Needless to say, those exist and
can be also produced. The peaks of $\bar s s$ $\phi$-like states 
should be shifted in mass
upward by  $2m_s\approx 250-300 \, MeV$,
but their contribution to dilepton spectra is proportional to
square of the electric charge $q_s^2=1/9$, which is 3 times smaller than
the average of  $q_u^2=4/9,q_d^2=1/9$. Strange states are more
promising for pseudoscalar/scalar signals to which we turn below.

\section{Generalities}
\label{sec_newqgp}
\subsection{From annihilation rates to the observed dilepton spectrum}
\label{sec_duality}
Before we proceed to realistic rates, based on lattice-based
quasiparticle interaction, let us explain few important points by
using
a much simpler model.

As explained above, we expect certain structures (``bumps'') to exist
in the annihilation rate, related to bound states as well as
near-threshold enhancement. Their exact shape is determined both
by the interaction between the annihilating quarks
(to be studied in detail in this work)
as well as the interaction with surrounding matter. The latter leads to a
``width'' which we will leave to be studied later elsewhere.

Because of this, it is instructive first to show how sensitive
is the resulting shape of the observable structures to such widths.
Let us illustrate the point by the following example.
As a representative of a (near-threshold) bump we start with the
rate in which the ``formfactor'' $F$ (\ref{ff}) is written as
\be F_{bump}/24=M_0{e^{-{(M-M_0)^2\over 2\Gamma^2}}
\over (2\pi)^{1/2} \Gamma}+{1 \over exp[-2*(M-M_0)]+1}\ee
In Fig.\ref{fig_3bumps}(a) one can see three examples of such bumps,
with different widths $\Gamma$.

  Although the width is different, the integral over the bump is kept
  the same. Furthermore, this integral is normalized to
``missing'' strength of the spectral density due to absent
annihilation rate between $M=0$ and $M=M_0$ (see the second term).
We did so because one has to be consistent with the so called
{\em quark-hadron  duality}, leading to conservation of the
total area of the spectral density.
Provided that one is using a sufficiently simple
ansatz\footnote{Otherwise just one sum rules does not have a
  predictive power, as experience of QCD sum rules had shown.},
as we do now, the duality restriction
provides a valuable relation between the bump strength and
the position of the threshold $M_0=2M_q$.

(Discussion of this and higher order duality relations
in vacuum correlators can be found in the classic paper
on the QCD sum rules \cite{SVZ}, for
more general
discussion of duality
see \cite{Shifman_duality} and for  finite T analogs of
it as well as Weinberg-like sum rules see
\cite{Kap_Shur}.)

  Here comes the main point we want to make in this section.
Although three bumps in Fig.\ref{fig_3bumps}(a)
look very different, those are for fixed $T$. In experiment we
have to integrate the rate over the expanding fireball: let us see
how these bumps will look after it is done.

   Although
the space-time evolution of the QGP phase at RHIC
is complicated hydro explosion, here and below we will
compare with the prediction for RHIC using the same parameterization
  as used by Rapp in  \cite{Rapp_RHIC},
concentrating only in the QGP and mixed phase.
The physical basis for such simplification is that the invariant mass of
the
dileptons is the same in all frames, and therefore motion of the
matter
can be ignored. (It would not be possible if e.g. one wants to calculate
the transverse momentum of dileptons, or other non-invariant
property.)

We sketch few details here for completeness.
The thermal rate is convoluted with the space time history by
\be
\frac{dN^{thermal}_{l^+l^-}}{dM}=\int_{\tau_0}^{\tau_f} d \tau V(\tau) \int 
d^3q \frac{M}{q_0}\frac{dR^{thermal}_{l^+l^-}}{d^4q}
\ee
The t-dependent volume is modeled as expanding cylinder
\be
V(\tau)=(z_0+v_zt) \pi  \lgroup r_0 +\frac{1}{2} a_{\bot} t^2 \rgroup ^2
\ee
where $r_0=6.5 fm$ is the initial transverse overlap, $z_0=0.6$ the initial
longitudinal lenght, $v_z=1.4c$ is the relative 
longitudinal expansion
velocity of the fireball edges and $ a_{\bot}=0.035 c^2/fm $ is such that at 
typical freeze-out time
$t_{f}=15-20 fm/c.$ we get a final transverse velocity $v_{\bot}=0.6c$.
The initial temperature of the system is set to $T=370 MeV$. The
expansion is considered is-entropic with
the entropy density of the QGP phase  given by
\be
s_{QGP}=(16+10.5 N_f) \frac{4\pi^2}{90}T^3
\ee
This fixes the time dependence of the temperature, leading to the transition 
at $t\approx=4 fm/c$. The
mixed phase takes place between $4 fm/c$ and $9 fm$. The fraction of QGP in 
the plasma phase is also
calculated from standard requirement of constant entropy
\be
S_{tot}/V(\tau)=  s_{QGP}(T_c) f_{QGP} +s_{HG}(T_c)(1-f_{QGP})
\ee

In doing the convolution, one has to define how the parameters of
the rate,
$\Gamma,M_0$ depend on $T$.  The results shown in
Fig.\ref{fig_3bumps}(b)
correspond to $\Gamma$ being independent on $T$ while the mass
depending linearly on it, $M_0=6T$.
One can see from a comparison of the input to
output figures that three cases, although still distinct, look much
more similar. The reason for that is that we kept the integral over
the spectral density constant, conforming to parton-hadron duality.
The sensitivity to (unknown) width of the bump is reduced
because of the averaging.

The effect of time averaging is maximal in this example because we took
a simple linear dependence for threshold $M_0=6T$. In reality,
the dependence is nonlinear with a minimum (see dash-dotted curve
in  Fig.\ref{fig_masses}). Although the precise shape and position
of zero binding point are not yet known (better lattice results, please...)
one may think that real dependence of $M_0$ on $T$ is weaker and
the effect of averaging be less significant.

\begin{figure}[t]
\centering
\includegraphics[width=6.cm]{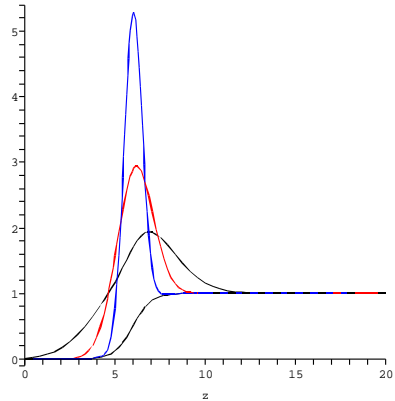}
\includegraphics[width=6.cm]{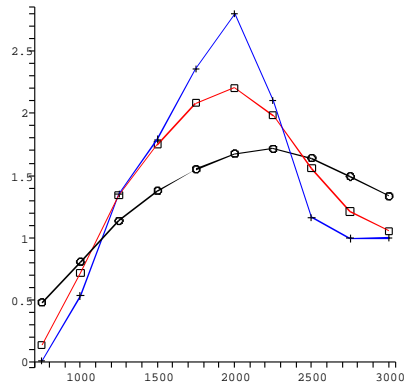}
   \caption{\label{fig_3bumps}
(a) An example of spectral densities, all of the satisfying the
duality condition, $before$ time integral. (b) This is how the corresponding
spectra
look like $after$ the averaging over the expanding fireball is
performed.
The lines marked by crosses, squares and circles correspond to most
narrow, the middle and the widest peaks in (a), respectively.
}
\end{figure}

\subsection{Non-relativistic approach and Green functions}

As mentioned in the Introduction, the main goal of this paper is to
find out how the strong interaction
among the initial quarks modifies  the dilepton production.
Such calculations simplify significantly if one can use
the non-relativistic approach. This is clearly possible
for invariant masses close to threshold.

The modification is the same in the inverse process, and
annihilation of $e^+ e^- $
pairs into $\bar q q$ has been studied in detail in
leptonic colliders. The nonrelativistic methods we use benefit in
particular from discussion of Coulomb enhancement of
$t \bar{t}$ pairs near threshold in \cite{FK_88},\cite{Strassler:1990nw}.

In \cite{Strassler:1990nw} the cross section $\sigma(e^+e^ \rightarrow t 
\bar{t})$ is analyzed to
leading-logarithmic order in QCD in the {\it non relativistic limit} (close 
to threshold). Their main
result, already used in \cite{FK_88}, is the modification of the threshold 
factors from the
well known leading order expression

\be
\sigma_{LO}=\frac{4\pi\alpha_{QED}^2 e_t^2 }{3s} N_c 
\sqrt{(1-\frac{4m_t^2}{s})} (1+\frac{2m_t^2}{s})
\ee
to
\be
\label{sigmaG}
\sigma=\frac{4\pi\alpha_{QED}^2 e_t^2 }{3s} N_c \frac{24 \pi 
\Im{G_{E+i\Gamma_t}(0,0)}}{s}
\ee
Where E is the center of mass energy and $\Gamma_t$ is the width of
the
top quark (related to lifetime due to weak decays).
The relation for dilepton production rate is obtained by
obvious substitution of $\Im{G}/s$ into $F_Q$ in (\ref{ff}).

As a further clarification,  in the limit $\Gamma=0$ we can rewrite
\be
\label{spectrum}
\Im G_{E+i0^+}(0,0)=\sum \psi_n(0) \psi_n(0)^{ \dagger} \delta(E_n-E)
\ee
Note that the sum runs over $all$ states, including scattering states, where 
the sum should be replaced
by an integral. This expression explains quite clearly
the connection between  the Green\'{}s function
and the standard non relativistic formula for the annihilation of bound 
states (\ref{spectrum})
gives the flux factor that multiplies the annihilation cross section at zero 
momentum
\be
\sigma=\sigma_{q\bar{q}}(s=4m_q)\arrowvert \psi_n (0) \arrowvert ^2
\ee

  The non-relativistic Green\'{}s  function
$G_{E+i\Gamma_t}(r,\bar{r})$  obeys
the usual  Schrodinger equation:
\be
\label{SGF}
[-\frac{1}{m}\vec{\nabla}^2 + 
V(\vec{r})-(E+i\Gamma)]G_{E+i\Gamma}(r,\bar{r})=\delta^{3}(\vec{r}-\vec{\bar{r}})
\ee
with  inter-particle potential
$V(\vec{r})$. Analytic expressions for  Green\'{}s function
for the pure Coulomb potential are well known,
see e.g. \cite{FK_88}. For realistic
lattice-based potential we will use  a numerical method following
\cite{Strassler:1990nw}, which is
valid for all potentials less singular than $1/r^2$.
The benefit of it is that one avoid summations over levels, and
bound states are automatically included together with scattering ones.
As a test, we checked that it reproduces well known results for
Coulomb potential with accuracy at least few per mill.

The Green\'{}s function, thus, contains, at the non relativistic level, all 
the information
needed about the states of the two particle system. This
is important because we are interested not only in bound states but
also in
modification of scattering states, especially
close to zero energy, to be responsible for the threshold effects we
will
study.
The cross section of $q+\bar{q} \rightarrow e^+ +e^-$ is calculated by 
simply inverting
the previous one from phase space considerations (neglecting the lepton 
masses)
\be
\sigma_{q+\bar{q} \rightarrow l^+ + l^-}(s)=\frac{1}{N_c^2}\frac{\sigma_{ 
l^+ + l^-  \rightarrow  q+\bar{q} }(s)}
{(1-\frac{4m_q^2}{s})}
\ee
\subsection{Lattice-based  potentials for static quarks}

As in \cite{SZ_bound} we use a potential
extracted by parameterizing the Bielefeld data \cite{potentials} on the 
effective static potentials
(for details see \cite{SZ_bound}). These potentials are used to solve the 
Schrodinger (or Klein Gordon)
equation for different temperatures.

Another ingredient taken from lattice calculation is the quasiparicles 
masses. In \cite{masses}  Petreczky et al.
have shown that the masses of the quarks in the region of $1.5-3 Tc$ stay 
roughly constant at a value of $Mq \approx 1 GeV$
As in \cite{BLRS} we fix this value all the way up to Tc, being aware that 
the quasiparticles should become heavier
close to the transition temperature.

It was also pointed out in \cite{BLRS} that in order to trigger the phase 
transition at $T=T_c$ the lowest $q \bar{q}$
bound state is required to become massless. The previous static potential is 
not enough to achieve this, and a quasi-local
interaction induced y the {\it instanton-antiinstanton molecules} 
\cite{BLRS} \cite{SZ_bound} is needed. 
We will model this interaction by the following 
potential:
\be
V_{inst}=\frac{U_0}{(r^2+\rho^2)^3}
\ee
where $\rho=1/3 fm$ is the typical instanton size. The value $U_0$ is fixed 
at $T_c$ by the requirement of
the lowest level to be exactly massless
(in the chiral limit). At higher temperature, the instantons are 
suppressed, which we
model with the following damping factor
\be
n(\rho,T)=n(\rho,T_c)\exp(-\frac{1}{3}(2N_c+N_f)(\pi \rho )^2 (T^2-{T^2}_c))
\ee
Finally, the polarization of the instanton molecules along the (Euclidean) 
time direction leads to a 
distinction of the $\pi$ and  $\rho$ channels, see discussion in \cite{BLRS}. 
To take this into account, the authors of \cite{BLRS} have  multiplied
the strength of the non-local interaction by a correction factor
$U_0\rightarrow U_0 F$ where for vectors $F=0.7 $.

\section{Results}

  First, let us explain our units.
 Since  the effective mass of quark quasiparticle in the temperature
interval considered is not known accurately, we use it as our basic
energy unit. In plots  twice this value appear as a threshold,
to which we  tentatively ascribe to it a particularly simple value
$2M_q=2\, GeV$: the reader should however be warned that this is
a ``GeV'' in quotation marks, to be rescaled appropriately later when
the value of quark effective masses in QGP will be better known.

\subsection{A  near-threshold region}
Although the underlying potentials describing interaction between
quasiparticles in the channels we consider are always attractive,
due to presence of bound/virtual states 
the effective scattering amplitude  is complex and
its real part may even change sign. In particular,
it is well known from quantum mechanics
that since a scattering amplitude changes sign
at energy equal to the position of the level,  an overall
attractive potential can cause 
effective interaction of positive energy particles
to be repulsive, if the level is close to zero,
 However this effect holds only for scattering states, or particles
 moved far 
away from the origin (where the potential is negligible). 
We however are interested in a shape of the annihilation signal:
and thus we are only concerned about the wave function at the origin.
Even in a case when
a level approach zero,
 the effect of the attractive potential is only to increase the
annihilation rate: there cannot be any opposite effect.

 All of it  can be studied in the simplest problem possible,
that of a spherical potential well:

\begin{eqnarray}
\label{well}
V= \cases{  0, &($r>a$)\cr
-V_0, &($r<a$)\cr}
\end{eqnarray}

Parametrizing $V_0=\frac{(2n+1)^2 \pi^2}{4 a^2 M}$ we find levels at 
threshold
($E=0$) for $n$ integer and $n$ is the number of levels below threshold. It 
is straightforward now to solve the Green\'{}s function equation for this 
potential (\ref{SGF}). Again, as we are only interested in the Green\'{}s 
function at
the origin, we only need to calculate the s wave equation with the standard
boundary conditions. The result of this simple calculation (in the limit
$\Gamma \rightarrow 0^+ $ is

\be
\label{GF_well}
\Im G_{E+i0^+}(0,0)=-\frac{1}{4}\frac{M\omega^2 k}{\pi((\omega^2-k^2)
\cos^2(\omega a)+k^2)}
\ee

And $\omega^2=M(E+V_0)$, $k^2=M E$. When $n$ is not an integer, the 
Green\'{}s
function close to zero behaves as $-\frac{1}{4}\frac{M k}{\pi}$, that is, 
the value of the free Green\'{}s function. When $n$ is integer, we have a 
level at $E=0$ what leads to a divergence. We see then that the 
modifications at threshold
come from the presence of bound states and virtual levels close to zero binding. In the case of
coulomb potential, the finite value of this object at threshold is a
consequence of the infinite number of states close to that energy.

We also observe that, according to (\ref{sigmaG}), the cross section at any
energy is bigger (or equal) than the free one, as correspond to an attractive potential.
 It is interesting to note that
the oscillations at positive energy correspond to {\it virtual levels}, as the 
continuation of the denominator of \ref{sigmaG} to imaginary values of $k$ 
is the condition for bound states in the potential.  The position of the 
maxima, then, reflects the position of these virtual levels.

Repeating the calculation that lead to (\ref{impi}) (see, for 
example,\cite{RW}) with the cross sections expressed
in terms of the Green\'{}s function (\ref{sigmaG}) we obtain the following 
expression

\be
\label{ImPGF}
\Im{\Pi_{em}(M)}=-4\Im{G_{M-2m_q+i\Gamma_q}(0,0)}
\ee

The calculation of the modification of the spectral density is now straight 
forward. The parametrization of lattice
results plus the non local interaction described in the previous section 
gives us a close expression for the
inter-particle potentials at different temperatures that we can use to solve 
(\ref{SGF}). We do this numerically
by following the  numerical method of \cite{Strassler:1990nw} based
on finding two independent solutions of Schreodinger equations.
 We 
evaluate $\Im{G_{M-2m_q+i\Gamma_q}(0,0)}$ directly for ``realistic''
lattice-based potentials.

\begin{figure}[t]
\centering
\includegraphics[width=6.cm,angle=0]{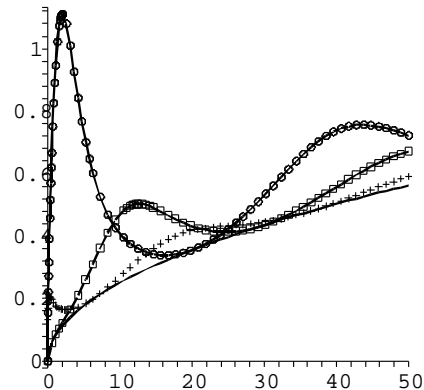}
   \caption{\label{fig_ImGf}
$\Im G_{E+i0^+}(0,0)$ versus energy/mass  for a=1 also in inverse 
mass units: Free case (solid), n=0.01 (cross), n=0.5 (box), n=0.9 (circle)}
\end{figure}

We still have to address the issue of the width. It is clear that the 
quasiparticles will have a thermal
width in the equilibrated medium. Perturbative calculations for this 
quantity  exist in the literature
and to lowest order they are
(at rest)
\be
\Gamma=a(N,N_f) \frac{g_s^2TC_F}{4\pi}
\ee

With $a(N,Nf)\approx 1.40$ for $N=3$ and $N_f=2$. For $T\approx (1-3)T_c$ 
and $\alpha_s\approx 0.5$
gives a width $\Gamma \approx 0.4-1 GeV$. These values of the 
width, comparable to the quasiparticle
mass, are too big and would make very complicated to be able to extract a 
value for the masses from
lattice data. Of course, there is no reason to believe than the perturbative 
expression makes
sense for such a big value of $\alpha_s$. That is why we follow a 
phenomenological approach, leaving the
width as a free parameter and presenting our calculations for different 
values.

\begin{figure}[t]
\centering
\includegraphics[width=6.cm,angle=-0]{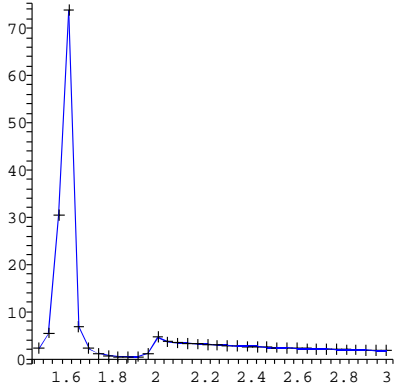}
   \caption{\label{ImPiTc}
 Modification of the spectral density at $T_c$ with respect to invariant mass in 
units of $M_q$.
 The peak at 1600 MeV corresponds to the $\rho$ (non relativistic approximation). 
The peak at 2000 MeV reflects the threshold enhancement
}
\end{figure}

 We now move to present the results, starting
 with the worning.
 In figure \ref{ImPiTc}  we show the
modification of the spectral function, (the ratio (\ref{ImPGF}) over 
(\ref{ImPipert})) at $T=T_c$ ($\Gamma=10 MeV$)
. One can see the threshold
enhancement at 2 GeV and the peak around 1600 MeV corresponding to the bound
state that should represent 
 $\rho$ at $T=T_c$. 
Although the potential we used has the same suppression of instanton
molecules
as in \cite{BLRS}, instead of relativistic  Klein Gordon equation
used in that work we now use the nonrelativistic  Schrodinger equation 
and thus found much less deeply bound state, with only about 400 MeV binding.
(We checked that
 Klein Gordon equation with the same potential modified by 
for Schrodinger and by simulating
the velocity-velocity term by effectively doubling the coupling \cite{BLRS} 
\cite{SZ_bound} indeed shifts
$\rho$ much lower, to  about $600 MeV$, as found in that paper.) 

Thus at $T\approx T_c$
 the relativistic 
effects are very  important.
It means that one should either (i) develop fully relativistic theory
of Green functions and dilepton rates, or (ii) restrict the discussion to
the nonrelativistic domain at somewhat higher $T$ not too close to $T_c$. 
In this work we only follow the latter option: the reader 
thus should be aware 
 that our results should not be trusted close to $T_c$.

  What we learned  from these calculations is how the
spectral density changes as  the
bound states become less and less bound until the system arrives to 
zero binding (zero binding point).
 In figure 
\ref{ImPiofT}  we show the shape of the dilepton spectral density
 for different
temperature ($1.2$ $T_c$, $1.4$ $T_c$, $1.7$ $T_c$, $3$ $T_c$). 
One can observe how exactly enhancement in the bound state and 
threshold region changes. 
The hight of this enhancement depends on the proximity of the level; at $1.7$ $ 
T_c$, where the bound level is close to threshold, we observe a big modification of
the spectral function at $2GeV$. Note that  proximity of the level to 
threshold  happens in rather wide $T$ interval, roughly
between $1.5$ $T_c$ to $2$ $T_c$ (the zero
binding point \cite{SZ_rethinking}. Note also that
the enhancement is still seen at 
temperatures as high as $3$ $T_c$, when all  bound states
have already been melted.

We show in figure (\ref{ImPiofG})  the dependence of the modification 
factor with the width of the quasiparticles at a fixed temperature of $2$ 
$T_c$. The general features of the modification are the same.  However, the 
extra enhancement due to the proximity to threshold
is very sensitive to the width of the quasiparticles. But, we expect at 
least a factor 2 from the final calculation
of the rates with the standard expression (\ref{ImPipert}) for any 
reasonable value of the width.

\subsection{Dilepton signal at RHIC}
\label{sec_rhic_signal}

In this section we
 discuss the time-integrated dilepton production
rates. We start with the QGP phase.
We can observe an enhancement in the region close to $2$ 
GeV.
This is because in the whole region of temperatures accessible at RHIC we 
observe
the bump in the cross section due to threshold effect. The
contribution below threshold is strongly suppressed due to the mass effect. 
The
value observed comes from the bound state moving from threshold till the value 
at $T_c$.

When we look at the contribution from the mixed phase, we can observe a 
significant threshold contribution.
However the main contribution is given by the bound state at $T_c$. We remark again 
that this contribution should not
be considered too seriously due to the relativistic character of the bound 
states.

\begin{figure}[t]
\centering
\includegraphics[width=6.cm,angle=-0]{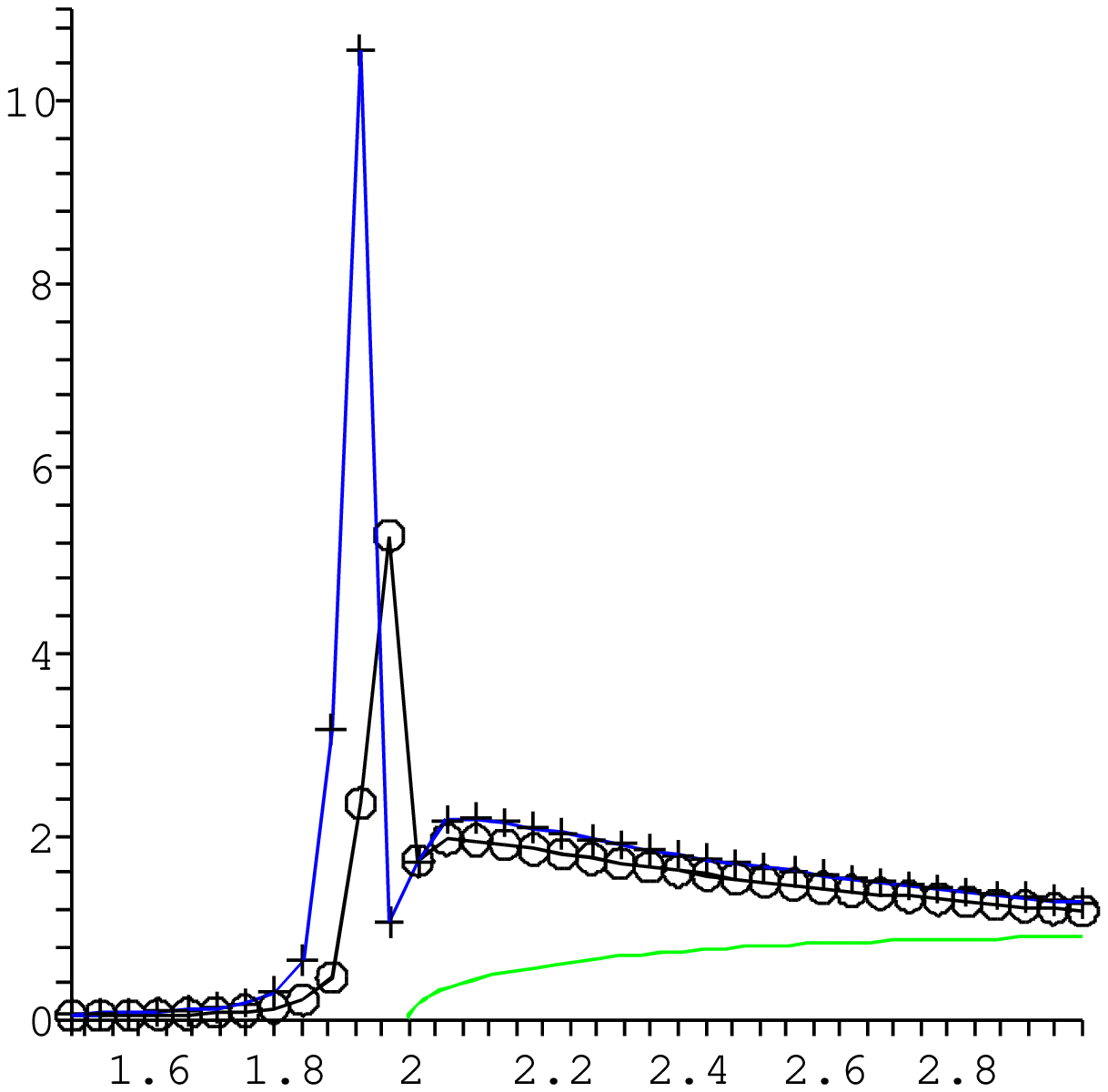}
\includegraphics[width=6.cm,angle=-0]{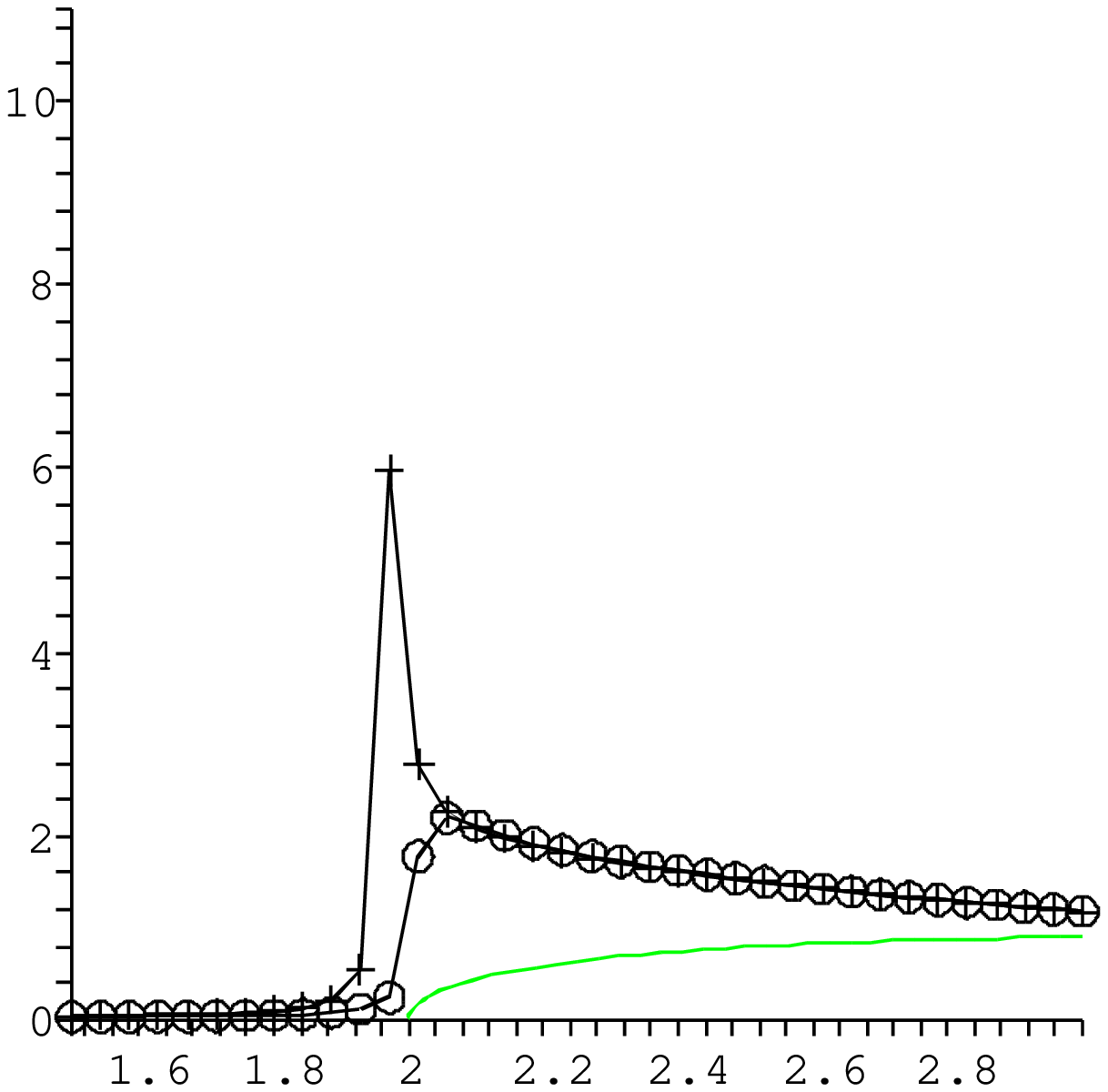}
   \caption{\label{ImPiofT}
 Modification of the spectral density versus the invariant mass in $M_q$ units  for different
temperatures: (a) 1.2 Tc (cross) and  1.4 Tc (circle), (b) 1.7 Tc (cross) and 3 Tc (circle) and the 
correction due to quark mass (line).
 }
\end{figure}


\begin{figure}[t]
\centering
\includegraphics[width=6.cm,angle=-0]{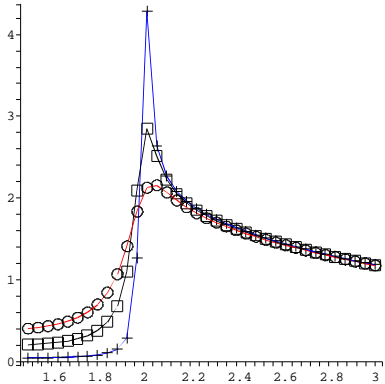}
   \caption{\label{ImPiofG}
 Modification of the spectral density versus the invariant mass in $M_q$ units .
Dependence  on the width  (at $T=2T_c$); $\Gamma=10 MeV$ (cross),
$\Gamma=50 MeV$ (box), $\Gamma=100 MeV$ (circle)   
 }
\end{figure}

Finally we show the total modification of the production due to the QGP. We 
show the effect of the widths. As the contribution of the mixed phase is much 
smaller than the QGP one, the effect of the bound states is not as 
prominent.
At the same time the contribution from the threshold produces an enhancement
that survives the time integration. 
Although both effects depend on the assume widths, they are present for values 
as big as $\Gamma=100 MeV$. Thus, 
 we do expect
an observable modification of the spectrum in sQGP, relative
  to the wQGP calculation by Rapp \cite{Rapp_RHIC}.

\begin{figure}[t]
\centering
\includegraphics[width=6.cm,angle=0]{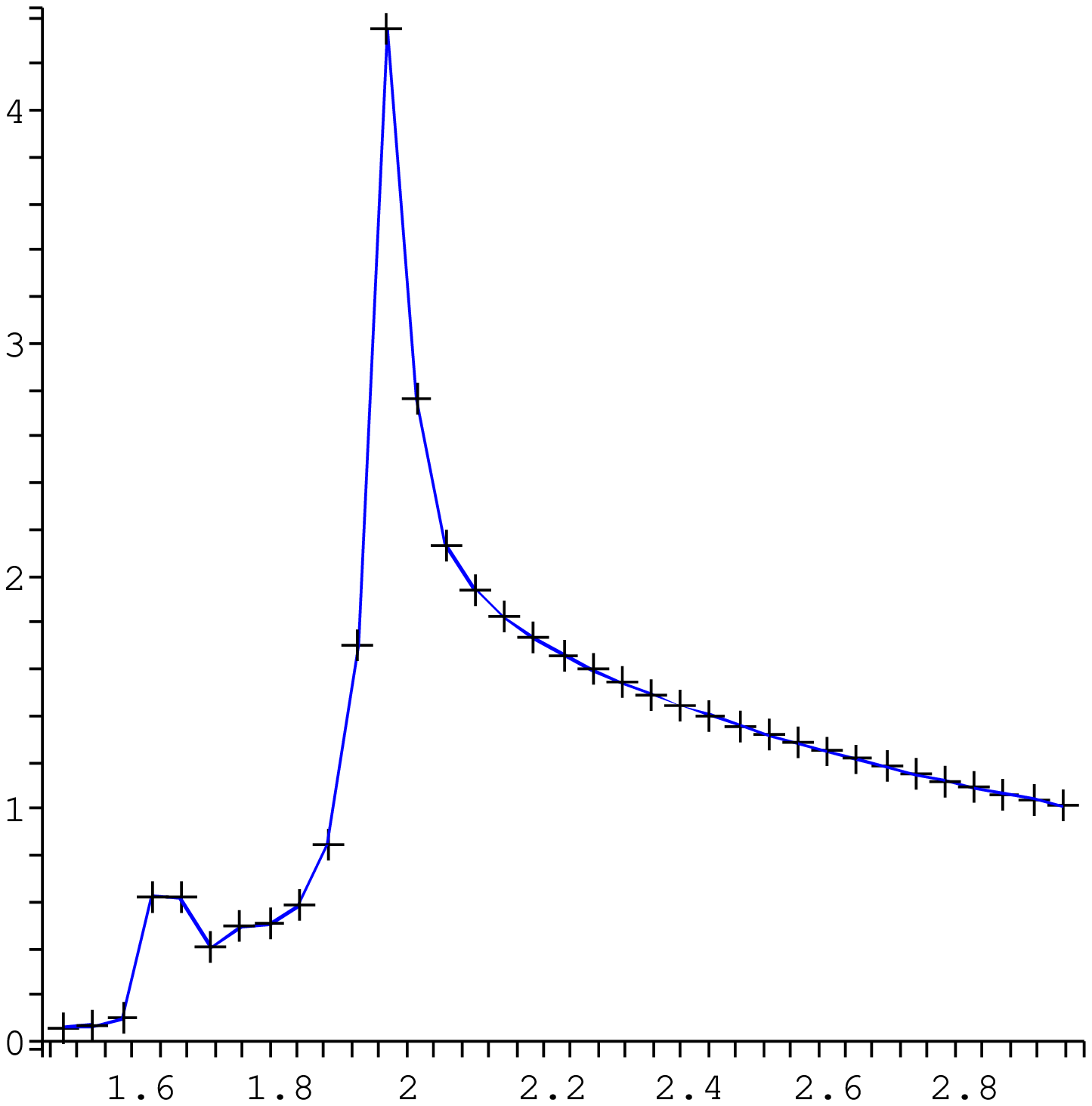}
\includegraphics[width=6.cm,angle=0]{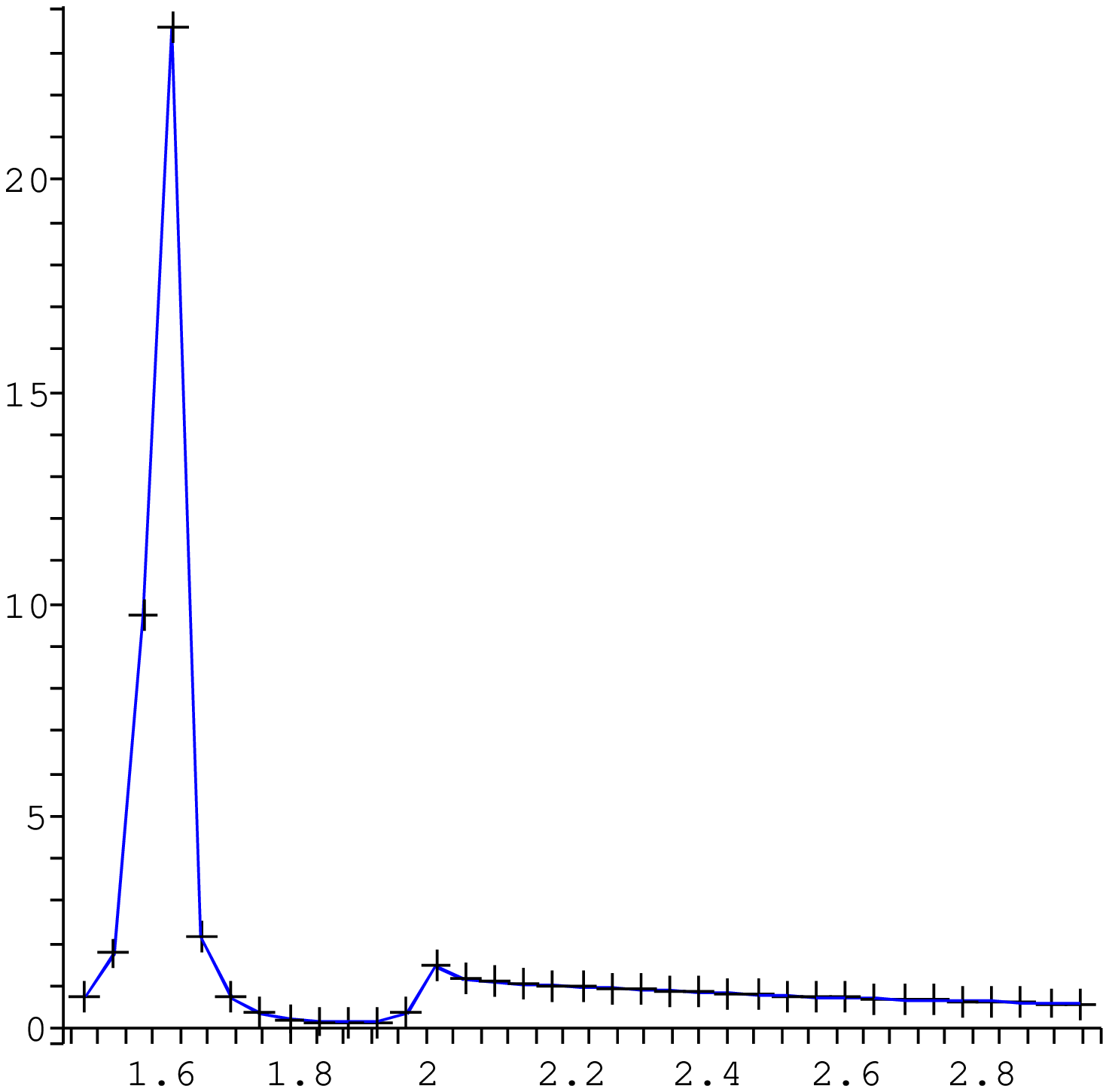}
   \caption{\label{fig_LTcomp}
 The  contribution
of (a) the QGP phase and (b) the mixed phase at RHIC to total
dilepton yield, normalized to the ``standard rates'' (note that both approach
 1 at high masses). The calculation is done for 
$\Gamma=10 MeV)$, the invariant mass is given
 in units of $M_q$.
  }
\end{figure}

\section{Two-photon signals}
 \label{sec_gammas}

At first glance, $\gamma\gamma$ spectroscopy looks even more promising than
dileptons, since such
 decays can happen for wider range of channels --
such as pseudoscalars, scalars and tensors.

\begin{figure}[t]
\centering
\includegraphics[width=6.cm,angle=0]{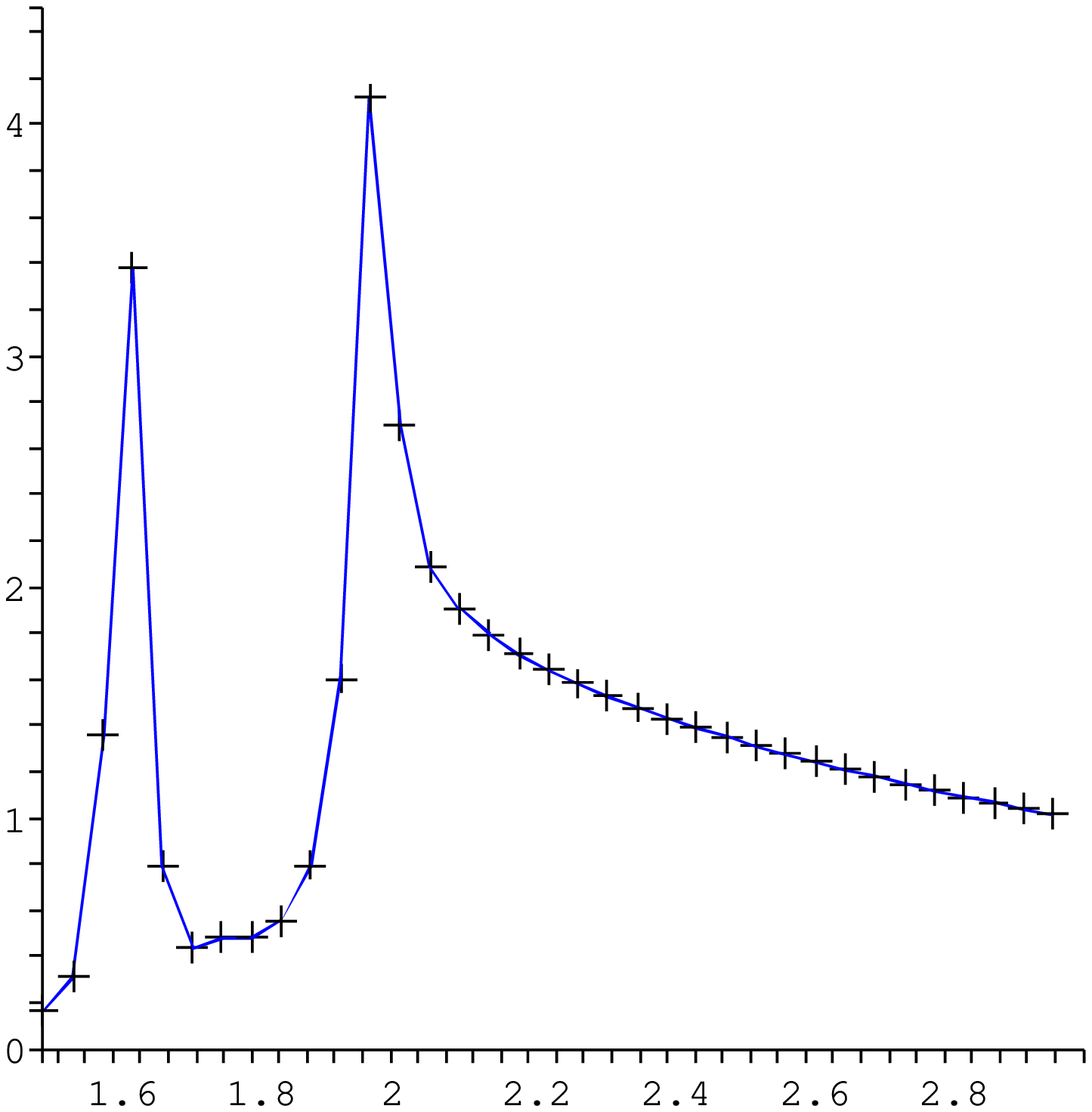}
\includegraphics[width=6.cm,angle=0]{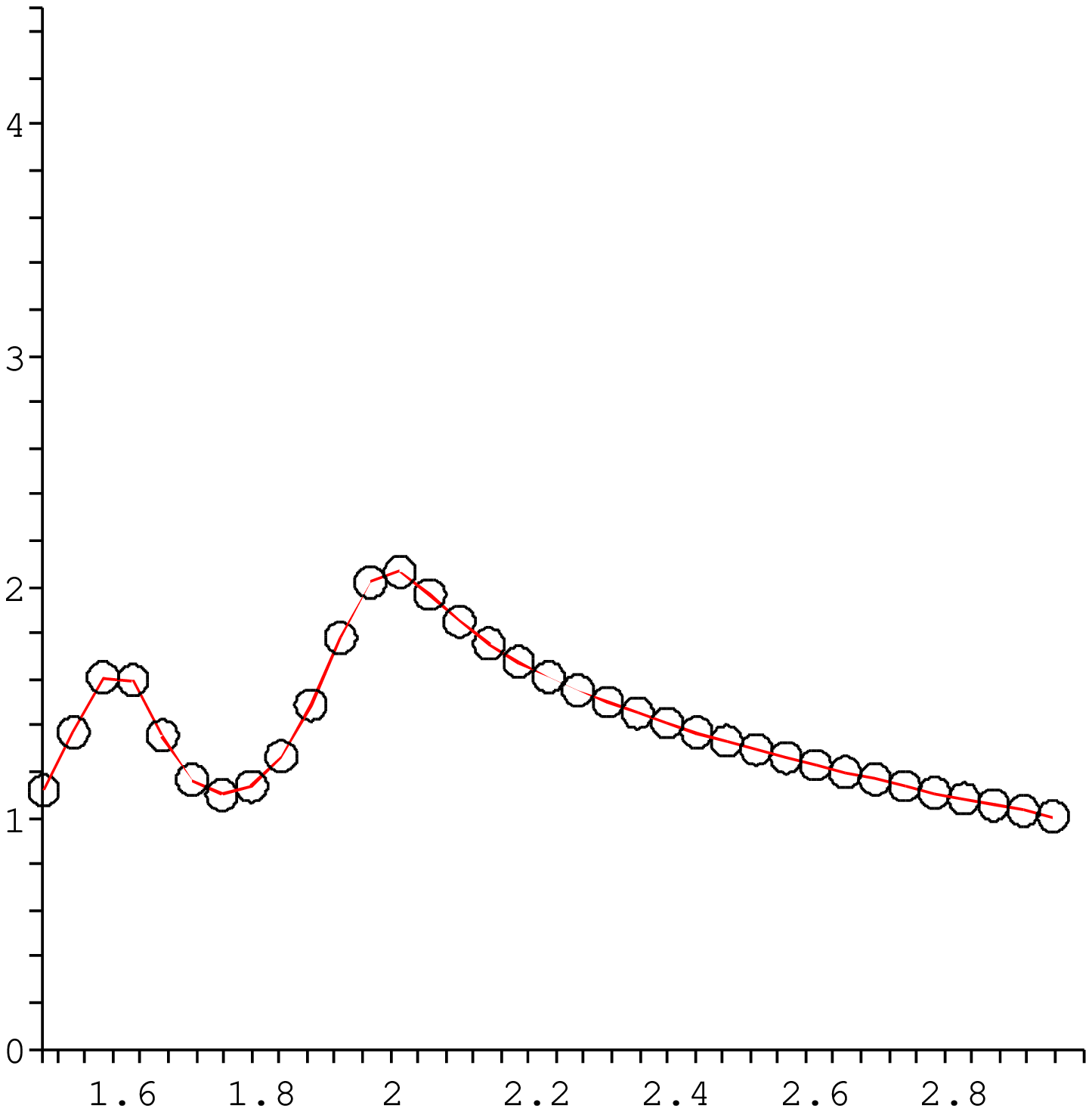}
   \caption{\label{fig_LTcomp}
The ratio of dilepton production from the QGP+mixed phase to 
``standarad rate'', vs the invariant mass 
(in $M_q$ units). The effect is shown for two widths (per quark),
(a) $\Gamma=10 MeV$, (b)
 $\Gamma=100 MeV$
  }
\end{figure}

{\bf Pseudoscalars}: In vacuum, as well as 
 the hadronic phase $T<T_c$ the pseudoscalars $\pi,\eta$
are Goldstone bosons, and their masses are protected by chiral
 symmetry,
and thus if all quarks be massless they should remain zero.
As it is wel known, the near-SU(3)-singlet  $ \eta'$ is not a Goldstone boson
due to chiral anomaly and instantons, which  explicitly violates the
chiral $U(1)$ symmetry. A sketch of levels
motion with $T$ is shown in Fig.\ref{fig_masses_PS}.
As $T$ grows, the mixing between 
$\eta'$ and $\eta$ is expected to change as well,
they start at $T=0$ close to pure SU(3) singlet
and octet, but above $T_c$ they eventually rotate
into purely light $\bar u u,\bar d d$  and strange $\bar s s$ states.
This is related with the fact that the role of instanton-induced
interaction deminishes with growing $T$, both because of reduced
wave function at the origin of the bound states and because of
instanton suppression by Debye screening.


For estimates of the rate we will need the corresponding
electromagnetic
widths, given by their total widths and diphoton branching ratios
from the Particle Data Tables:
\be \Gamma_{\eta\rightarrow\gamma\gamma}=1.18 \, keV* .39=.46\, keV \ee
\be \Gamma_{\eta'\rightarrow\gamma\gamma}=.2 \,MeV* 0.02=4 \,keV \ee

{\bf Scalars}: As it is well known,
the situation with scalar mesons is rather complex.
The debates of whether the lightest scalars such as $f_0(600)$
(known formely as $\sigma$ of Gel-Mann-Levy sigma model) is a $\bar q
q$ state go back decades, and some heavier
 scalars 
 are now believed to be partly diquark-antidiquark states, and also mix with
 the
 scalar
 glueball. Not going into this discussion we only comment that above
 $T_c$
(in the chiral limit) there must be degeneracy between pseudoscalar
 and scalar chiral partners 
 of the pion and eta mesons. In other words, if quark chirality
is conserved, there are identical
states made of $\bar q_L q_R$ and  $\bar q_R q_L$ quarks.

{\bf Tensors}:  
 $f_2(1270)$ is a famous example of a state seen in 
$\gamma\gamma$ collisions, its partial width is comparable to that
of scalars and pseudoscalars, namely
\be \Gamma_{f\rightarrow\gamma\gamma}=185
\, MeV* 1.4 10^{-5}= 2.6\, keV \ee
However at $T>T_c$ the situation is radically different:
the calculations made in \cite{SZ_bound}
show that all\footnote{The only exception is the scalar glueball
 channel in which color Casimir is factor 9/4 stronger and P-wave
 state
survives for a while. However glueball states are not relevant for
 electromagnetic
probes anyway.
} P-wave states seem to be dissolving very close to $T_c$,
and in fact the lattice-based effective potentials lead to a single
s-wave bound states.

An estimate of the production rate may be done as follows 
\be N_{\gamma\gamma}= \int \Gamma_{\gamma\gamma} n dt dV \ee
where the integral is done over space-time occupied by sQGP.
Using for estimate $t_{QGP}\sim 5 \,fm\sim 1/(40\, MeV)$ times gives small
parameter $\Gamma t_{QGP}\sim (4 \,keV/\,40 MeV)\sim 10^{-4}$, 
originating ultimately from the electromahnetic coupling $\alpha_{em}^2$. 
The integral of density over volume
 ndV is dominated by later stages and thus is approximately equal
to the particle number produced at chemical freezeout.  Thus
we conclude that for $\eta,\eta'$ 2-photon signal
the in-matter (modified) signals are about 4 orders of magnitude below
those coming after freezeout.

  If instead of $T\approx T_c$ we think now about the
contribution to the time integral of the early times, at the vicinity 
of the zero binding point, the temperature is about twice higher
 but the mass of the state we would like to produce, with the energy
$\approx 2M_q\approx 2\, GeV$ is also twice higher than the mass of
 the
$\eta'$, resulting in about the same Boltzmann factor.

Good news about these rates is that there should be more states
similar to $\pi,\eta$:
adding  scalars and states containing ``plasminos'' instead of quark
quasiparticles leads to additional factor 2*4=8 in the rate.

 Unfortunately there are also small parameters involved, leading to
further troubles for the   
 $\gamma\gamma$ signal. The widths  $\Gamma_{\gamma\gamma}$
of the pseudoscalar mesons in vacuum are proportional to $f_\pi,f_\eta$ etc
which (as one of the order parameters of chiral symmetry)
 must vanish at $T>T_c$ (in the chiral limit).  
What it means is simply the statement that those particles are
bound state of $\bar L R$ or $\bar R L$ (where we mean left and right
 handedness of quarks) and in the chirally symmetric QGP phase
$L$ and $R$ are like  different quark flavors, which simply
cannot annihilate
each other as they are not antiparticle of each other.
 In order that  to become possible, one should admit
chirality flip in the process, or the first power of a current
quark  in the amplitude, resulting in 
additional suppression factor
\be
{\Gamma_{\gamma\gamma} (T>T_c) \over \Gamma_{\gamma\gamma}(T<T_c)}\sim
 ({m_s\over M})^2  \ee  
which reduces the effect by about an order of magnitude for strange
quarks (e.g. in $\eta.\eta'$) and much more so for the pion.

Although the authors are not qualified to tell what exactly the limits
of experimental detection of this effect is, it looks to be rather
difficult to measure these effects.

\begin{figure}[t]
\centering
\includegraphics[width=6.cm]{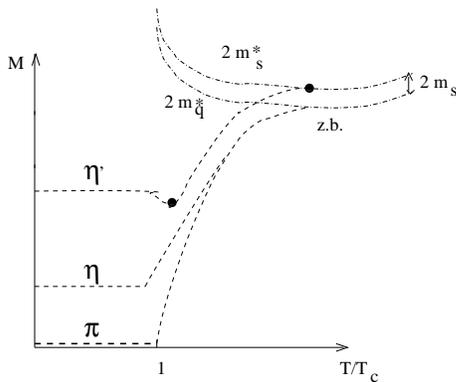}
   \caption{\label{fig_masses_PS}
Schematic  $T$-dependence of the masses of $\bar q q$
 pseudoscalar
states. 
Two dash-dotted lines shows a behavior of twice the quark quasiparticle
mass,
for light and strange quarks, respectively.
Two black dots indicate places where $\gamma\gamma$ signal may
be observable.
}
\end{figure}

\section{Summary and discussion}
We have proposed in this work new direction of dilepton experiments,
namely
looking for the bound vector $\bar q q$ states and threshold
enhancement in the QGP domain. We in particularly
identified
2 regions of the dilepton invariant mass
as most promising, corresponding to two endspoins of the sQGP domain.

In a nonrelativistic approach, we worked out a numerical method for
the determination
of the Green function and annihilation rate, using all states, bound
and unbound. We used realistic interaction and have studied in detail
the shape of spectral density as a bound level approach a threshold.

We have also performed realistic averaging over an expansing
fireball, using parameters for RHIC, and demonstrate that this
evraging does not eliminate the contribution of the peaks near
two ends of the sQGP region. The actual observability of states
still depend on
 the issue of the width of these states.
 We have decided at this point to leave this issue open for
further
studies. The reason for that is the  rather uncertain theory of sQGP and 
its near-perfect liquid properties. For the first investigations of
the ``photo-effect''-like reactions with splitting of the binary bound states
see \cite{SZ_ionization}. Lattice studies do see these peaks,
which is encouraging, but
the maximal entropy method they use in 
\cite{MEM} was not studied enough to tell how accurately the widths
can
really be estimated by it. 

We pointed out that in principle a dilepton study can be complemented
by
a $\gamma\gamma$ one,  for scalars and
pseudoscalars from sQGP. Interesting
signals can be expected around $T_c$ and the same threshold
phenomena near the endpoint as for dileptons. However, much higher
background makes it more difficult to do.

{\bf Acknowledgments.\,\,}
We thank G.Brown for useful discussions at the early stages of this
work,
and we are happy to thank V.Zakharov for an interesting discussion. 
This work was partially supported by the US-DOE grants DE-FG02-88ER40388
and DE-FG03-97ER4014.


\end{narrowtext}
\end{document}